\documentclass[letterpaper, 10pt, conference]{ieeeconf}

\IEEEoverridecommandlockouts                              

\usepackage{amsmath}
\usepackage{amsfonts, amssymb, amsbsy, mathtools, mathrsfs, nth, dsfont, bm}

\usepackage{array, multirow, booktabs, arydshln}

\usepackage{algorithm, algorithmic}

\usepackage{marvosym, textcomp, gensymb, accents}

\usepackage{stfloats, placeins, float}

\usepackage{verbatim, url, environ}

\usepackage{graphicx, epsfig, tikz, epstopdf, subcaption, xcolor}
\usetikzlibrary{shapes, arrows, calc, patterns, decorations.pathreplacing, backgrounds, positioning, fit}

\newtheorem{theorem}{Theorem}

\newtheorem{assumption}{Assumption}

\newtheorem{remark}{Remark}

\newcommand{\argmin}{\operatornamewithlimits{argmin}}

\allowdisplaybreaks[4]

\title{\LARGE \bf Quadratic-Programming-based Control of Multi-Robot Systems for Cooperative Object Transport}

\author{Si Wu$^{1}$, Zhengyan Qin$^{2}$, Tengfei Liu$^{1}$, and Zhong-Ping Jiang$^{3}$%
\thanks{This work was supported in part by Liaoning Revitalization Talents Program XLYC2203076, in part by NSFC grants 62325303 and 62333004, and in part by NSF grant ECCS-2210320.} 
\thanks{$^{1}$Si Wu and Tengfei Liu are with State Key Laboratory of Synthetical Automation for Process Industries, Northeastern University, Shenyang, 110819, China {\tt\footnotesize wusixstx@163.com; tfliu@mail.neu.edu.cn}.}
\thanks{$^{2}$Zhengyan Qin is with the Department of Mechanical Engineering, City University of Hong Kong, Hong Kong {\tt\footnotesize zhengqin@cityu.edu.hk}.}
\thanks{$^{3}$Zhong-Ping Jiang is with Department of Electrical and Computer Engineering, New York University, 360 Jay Street, Brooklyn, NY 11201, USA {\tt\footnotesize zjiang@nyu.edu}.}
\thanks{Corresponding author: T. Liu.}
\thanks{Digital Object Identifier (DOI): see top of this page.}}

\begin{document}

\maketitle
\thispagestyle{empty}
\pagestyle{empty}


\begin{abstract} 
This paper investigates the control problem of steering a group of spherical mobile robots to cooperatively transport a spherical object. By controlling the movements of the robots to exert appropriate contact (pushing) forces, it is desired that the object follows a velocity command. To solve the problem, we first treat the robots’ positions as virtual control inputs of the object, and propose a velocity-tracking controller based on quadratic programming (QP), enabling the robots to cooperatively generate desired contact forces while minimizing the sum of the contact-force magnitudes. Then, we design position-tracking controllers for the robots. By appropriately designing the objective function and the constraints for the QP, it is guaranteed that the QP admits a unique solution and the QP-based velocity-tracking controller is Lipschitz continuous. Finally, we consider the closed-loop system as an interconnection of two subsystems, corresponding to the velocity-tracking error of the object and the position-tracking error of the robots, and employ nonlinear small-gain techniques for stability analysis. The effectiveness of the proposed design is demonstrated through numerical simulations.
\end{abstract}


\section{Introduction}

Cooperative object transport serves as a benchmark scenario for demonstrating the coordination capabilities of robotic systems and has potential applications across various automated domains. Among different approaches, the caging strategy is one of the most common. In this setup, robots arrange themselves around an object to form a closure that constrains its motion, enabling them to push and guide it along a desired path \cite{Rimon-Blake-ICRA-1996}. 

Reference \cite{Wang-Takano-Hirata-Kosuge-IROS-2004} introduced a variable internal force control algorithm for three omnidirectional robots to cooperatively transport a cuboid object. Reference \cite{Brown-Jennings-IROS-1995} proposed a pusher-steerer framework that allowed more flexible interactions by relaxing the need for continuous contact with the object. Reference \cite{Spletzer-Das-Fierro-Taylor-Kumar-Ostrowski-IROS-2001} utilized vision-based methods to estimate inter-robot distances and orientations during transport. Reference \cite{Pereira-Campos-Kumar-IJRR-2004} developed a decentralized algorithm relying solely on local estimates of object orientation and neighbor positions. Building on these ideas, reference \cite{Fink-Michael-Kumar-RSS-2007} adapted collective transport for L-shaped objects following predefined trajectories, later extending the approach to environments with obstacles \cite{Fink-Hsieh-Kumar-ICRA-2008}.

Despite extensive technological and experimental advances in the fields of robotics and control engineering, systematic theoretical investigations into cooperative object transport remain scarce. The underlying theoretical challenges primarily arise from the uncertain dynamic coupling between multiple mobile robots and the object, as well as the requirement for real-time force distribution among the cooperating robots. The direct application of existing optimization algorithms for force allocation often introduces discontinuities and nonsmooth dynamics into the closed-loop system. The coexistence of such nonsmooth behaviors with complex dynamic couplings poses intrinsic difficulties to the stability analysis of the overall system.

In this paper, we investigate a typical scenario in which a multi-robot system cooperatively transports a spherical object. The object’s motion is driven by contact (pushing) forces exerted by spherical mobile robots. The object dynamics are modeled by a second-order integrator that captures the influence of the contact forces applied by the robots, while each robot is modeled by a first-order integrator with its velocity directly considered as the control input. The control objective is to ensure that the object follows a real-time velocity command through appropriate coordination of the robots’ motions.  

We propose a novel control approach based on quadratic programming (QP) to cooperative object transport. By treating the robots’ positions as virtual control inputs, the velocity-tracking problem of the object is formulated as a QP that assigns desired contact forces to the robots. This formulation enables the robots to collectively generate the desired contact forces required to drive the object along the target velocity while maintaining a closure around it. A dedicated velocity-tracking controller is designed for each robot. Particular attention is given to constructing the QP objective function and constraints to guarantee the feasibility and uniqueness of the solution. Furthermore, nonlinear small-gain techniques are employed to address the interaction between the object’s velocity-tracking error and the robots’ position-tracking errors. We prove that, under suitable choices of controller parameters, the cooperative transport objective can be achieved with the proposed design. For recent developments in QP-based control methods for constrained systems, see \cite{Ames-Coogan-Egerstedt-ECC-2019}, and for a survey of the nonlinear small-gain theorem, see \cite{Jiang-Teel-Praly-MCSS-1994}.

\section{Problem Formulation} \label{section_problemformulation}

This paper investigates a cooperative object transport system composed of multiple mobile robots and a single object to be transported. The object’s motion is driven by the contact forces exerted by the mobile robots. The objective is to ensure that the object follows a real-time velocity command by appropriately controlling the movements of the mobile robots.

\subsection{A Cooperative Object Transport System}

Consider $N$ spherical mobile robots with an identical and constant radius $D>0$. For each robot $i\in\mathcal{N}=\{1,\ldots,N\}$, the motion is described by
\begin{align}
\dot{p}_i&=v_i,\label{eq_plantmodel_integrator}
\end{align}
where $p_i\in\mathbb{R}^n$ and $v_i\in\mathbb{R}^n$ represent the position and the velocity of the robot's center, respectively.

The object to be transported is also spherical, with a constant radius $D_o>0$, and its motion is modeled as
\begin{subequations}\label{eq_objectmodel_doubleintegrator}
\begin{align}
\dot{p}_o&=v_o,\label{eq_objectmodel_doubleintegrator_objectkinematics}\\
\dot{v}_o&=f(p_1-p_o,\ldots,p_N-p_o),\label{eq_objectmodel_doubleintegrator_objectdynamics}
\end{align}
\end{subequations}
where $p_o\in\mathbb{R}^n$ and $v_o\in\mathbb{R}^n$ represent the position and the velocity of the object's center, respectively. The function $f:\mathbb{R}^n\backslash\{0\}\times\cdots\times\mathbb{R}^n\backslash\{0\}\to\mathbb{R}^n$ represents the influence of the net external force acting on the object, and is defined by
\begin{align}
&f(p_1-p_o,\ldots,p_N-p_o)\nonumber\\
&=-k_f\sum_{i=1}^{N}\max\{D+D_o-|p_i-p_o|,0\}\frac{p_i-p_o}{|p_i-p_o|}\label{eq_def_acceleration}
\end{align}
for $p_i\neq p_o$ for $i=1,\ldots,N$, where $k_f$ is a positive constant.

\begin{remark}
The system setup described in this paper aligns with established methods in the literature on cooperative object transport \cite{Howell-book-2001,Berman-Lindsey-Sakar-Kumar-Pratt-RSS-2010}. With the single-integrator model \eqref{eq_plantmodel_integrator}, it is assumed that the mobile robots are equipped with well-designed velocity controllers capable of adequately compensating the contact forces. In \eqref{eq_def_acceleration}, the term $f$ represents the repulsive contact forces between the robots and the object. The contact force exerted by robot $i$ becomes nonzero only when $|p_i-p_o|< D+D_o$. Its direction is aligned with the line connecting the centers of robot $i$ and the object, and its magnitude is proportional to $D+D_o-|p_i-p_o|$ with the proportionality constant $k_f$ determined by the mechanical properties of the system such as the object's mass and the elastic coefficients at the contact points. It should be noted that the proposed model neglects rotational dynamics and possible friction forces between the robots and the object, allowing us to concentrate on the fundamental challenges inherent in cooperative transport.
\end{remark}

\begin{figure}[h!]
\centering
\includegraphics[width=0.8\linewidth]{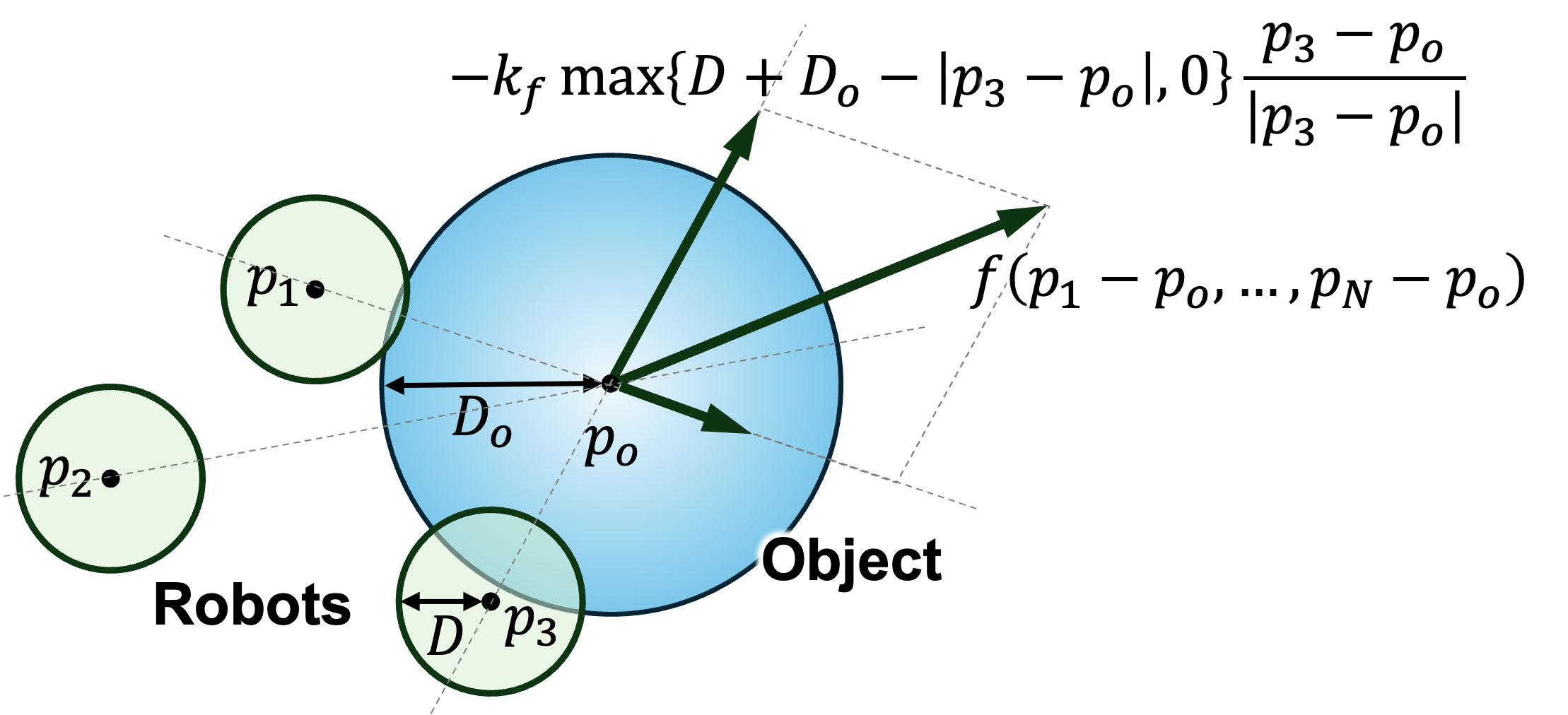}
\caption{Contact forces between the robots and the object.}
\label{figure_contactforce}
\end{figure}

\subsection{Control Objective and Assumptions}
\label{subsection_controlobjective}

\begin{figure}[h!]
\centering
\includegraphics[width=\linewidth]{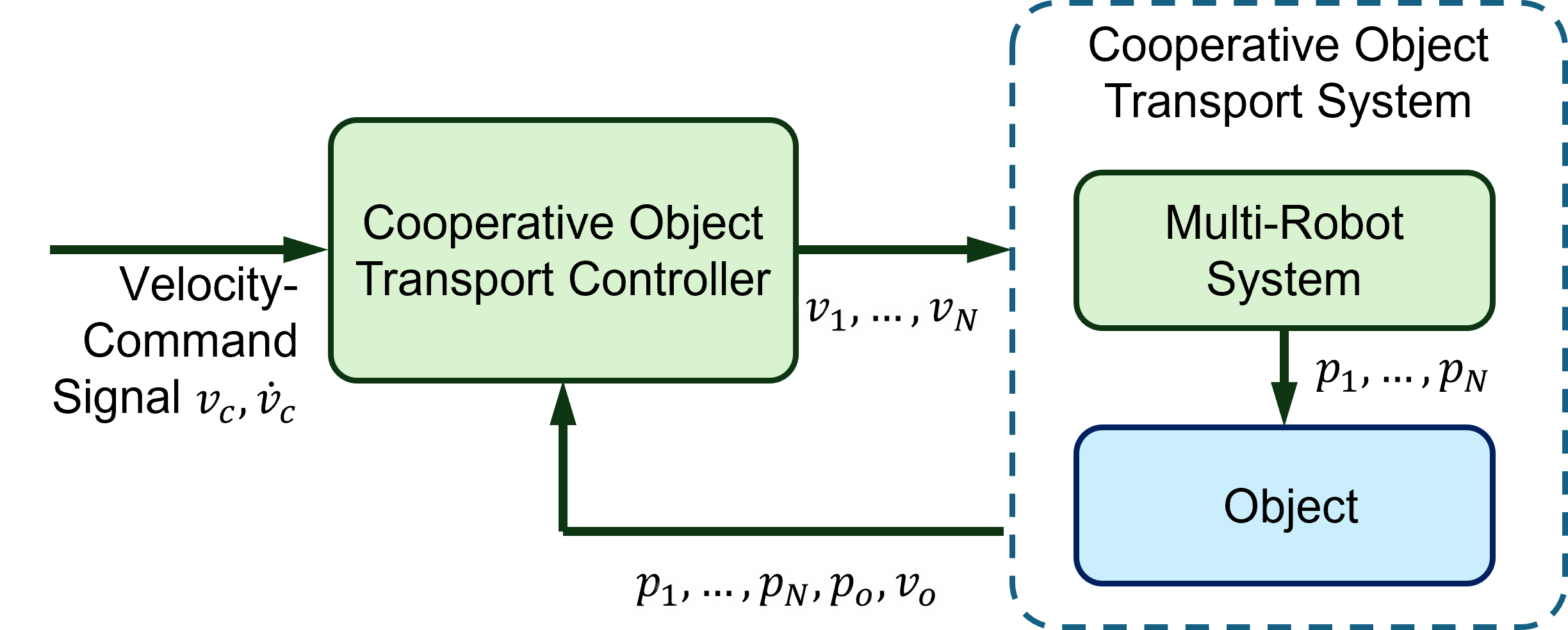}
\caption{Proposed control system structure of the cooperative object transport system.}
\label{figure_expected_control_diagram}
\end{figure}

Let $v_c$ denote the velocity-command signal for the object. Given that $v_1,\ldots,v_N$ are the control inputs of the robots, our objective is to design a controller of the form
\begin{align}
(v_1,\ldots,v_N)=\psi(p_1,\ldots,p_N,p_o,v_o,v_c,\dot{v}_c),\label{eq_def_controller_form}
\end{align}
such that, under specific initial conditions, the object is transported by the contact forces in such a way that its actual velocity $v_o$ practically follows the velocity-command signal $v_c$. In practical applications, $v_c$ is typically generated by a high-level planner or a supervisory controller. The desired structure of the overall control system is illustrated in Figure \ref{figure_expected_control_diagram}.

We impose smoothness and boundedness conditions on the velocity-command signal $v_c$.

\begin{assumption}\label{assumption_command}
The velocity-command signal $v_c:\mathbb{R}_+\to\mathbb{R}^{n}$ is smooth, and there exist positive constants $\bar{v}_c$, $\bar{v}_c^d$ and $\bar{v}_c^{dd}$ such that
\begin{align}
|v_c(t)| \le \bar{v}_c,\qquad |\dot{v}_c(t)| \le \bar{v}_c^d,\qquad |\ddot{v}_c(t)| \le \bar{v}_c^{dd}\label{eq_assumption_command_upperbound}
\end{align}
for all $t\geq 0$.
\end{assumption}

\section{Controller Design and Main Result} \label{section_controller_design_new}

In this section, we first consider the positions $p_i$ of the mobile robots as virtual control inputs and propose a QP-based controller for the object such that the velocity $v_o$ tracks the velocity-command signal $v_c$. Then, with the velocities $v_i$ as the control inputs, we design controllers for the mobile robots such that the positions $p_i$ tracks the ideal positions $p_i^*$ generated by the QP-based controller.

\subsection{Velocity-Tracking Control of the Object}
\label{subsection_objectvelocitytracking}

Consider the motion of the object. Given the dynamics described by \eqref{eq_objectmodel_doubleintegrator_objectdynamics}, we have
\begin{align}
\dot{v}_o-\dot{v}_c&=f(p_1-p_o,\ldots,p_N-p_o)-\dot{v}_c\nonumber\\
&=-k_f\sum_{i=1}^{N}\max\{D+D_o-|p_i-p_o|,0\}\frac{p_i-p_o}{|p_i-p_o|}\nonumber\\
&~~~-\dot{v}_c\label{eq_positionloop_objectvelocitytrackingerror}
\end{align}
We consider $p_i$ for $i=1,\ldots,N$ as the virtual control inputs, and use $p_i^*$ for $i=1,\ldots,N$ to represent the ideal virtual control inputs. Taking into account the characteristics of the contact forces between the robots and the objects, we propose the following QP-based virtual controller:
\begin{align}
p_i^*=p_o+l_i(D+D_o-s_i^*)\label{eq_velocityloop_virtualcontroller}
\end{align}
with
\begin{subequations}\label{eq_velocityloop_QP}
\begin{align}
&(s^*_1,\ldots, s^*_N)=\argmin_{(s_1,\ldots,s_N)} \left\{\epsilon \sum^N_{i=1}|s_i|^2\right. \notag \\
&\left.+\left|k_f\sum^N_{i=1}l_is_i-k_v(v_o-v_c)+\dot{v}_c\right|^2\right\} \\
&\text{s.t.}~s_i\ge 0,\qquad\text{for}~ i=1,\ldots, N,
\end{align}
\end{subequations}
where $\epsilon$ and $k_v$ are positive constants, and $l_1,\ldots,l_N\in\mathbb{R}^n$ are constant vectors. With $l_1,\ldots,l_N$ satisfying some conditions to be specified later, we can prove that the QP \eqref{eq_velocityloop_QP} is feasible and admits a unique solution. Then, we use $\phi_i:\mathbb{R}^n\times\mathbb{R}^n\to \mathbb{R}_+$ to denote the map from $(v_o-v_c,\dot{v}_c)$ to $s_i^*$ resulting from solving the QP, that is,
\begin{align}
s_i^*=\phi_i(v_o-v_c,\dot{v}_c).\label{eq_positionloop_virtualcontrollerfunction}
\end{align}

\begin{remark}
One may recognize that the objective function of the QP is motivated by finding $s_i^*$ such that $-k_f\sum^N_{i=1}l_is_i^*$ and thus $f(p_1^*-p_o,\ldots,p_N^*-p_o)$ (according to \eqref{eq_positionloop_objectvelocitytrackingerror} and \eqref{eq_def_acceleration}) closely approximate $-k_v(v_o-v_c)+\dot{v}_c$. In this manner, the velocity-tracking control objective for the object can be achieved in the ideal case that $p_i=p_i^*$ for $i=1,\ldots,N$. The additional term $\epsilon \sum^N_{i=1}|s_i|^2$ in the objective function serves to guarantee strict convexity, thereby ensuring the uniqueness of the QP solution \cite[Proposition 1.1.2]{Bertsekas-book-1997}. Furthermore, incorporating the $\epsilon \sum^N_{i=1}|s_i|^2$ term also minimizes the total magnitude of contact forces, and guides the robots to maintain a closure configuration around the object.
\end{remark}

\subsection{Position-Tracking Control of the Mobile Robots}
\label{subsection_robotspositiontracking}

Now, consider the position-tracking errors of the mobile robots. For each $i=1,\ldots,N$, with the ideal position $p_i^*$ generated by \eqref{eq_velocityloop_virtualcontroller}, we define the position-tracking error
\begin{align}
p_i-p_i^*=p_i-p_o-l_i(D+D_o-s_i^*)
\end{align}
with $s_i^*$ given by \eqref{eq_positionloop_virtualcontrollerfunction}. Suppose that $\phi_i$ is Lipschitz continuous (this is to be verified later). Then, we can use the Dini derivative $D^+s_i^*$ to represent the changing rate of $s_i^*$. See \cite{Giorgi-Komlosi-RMSES-1992} for the definition of the Dini derivatives. By also using \eqref{eq_plantmodel_integrator} and \eqref{eq_objectmodel_doubleintegrator_objectkinematics}, we have
\begin{align}
D^+(p_i-p_i^*)&=\dot{p}_i-\dot{p}_o+l_iD^+s_i^*\nonumber\\
&=v_i-v_o+l_iD^+s_i^*.
\end{align}
We consider $l_iD^+s_i^*$ as a perturbation term, and design a position-tracking controller
\begin{align}
v_i=-k_p(p_i-p_i^*)+v_o\label{eq_positioncontroller}
\end{align}
where $k_p$ is a positive constant.

\subsection{Main Result}

Theorem \ref{theorem_main} shows the effectiveness of the proposed design for the achievement of the control objective described in Section \ref{subsection_controlobjective}.

\begin{theorem}\label{theorem_main}
Consider the cooperative object transport system modeled by \eqref{eq_plantmodel_integrator}, \eqref{eq_objectmodel_doubleintegrator} and \eqref{eq_def_acceleration}, and the controller composed of \eqref{eq_velocityloop_virtualcontroller}, \eqref{eq_velocityloop_QP} and \eqref{eq_positioncontroller}. Choose $l_1,\ldots,l_N$ such that
\begin{itemize}
\item the vectors $l_1, \dots, l_N$ are unit vectors that positively span $\mathbb{R}^n$, i.e.,
\begin{subequations}
\begin{align}
|l_i| &= 1,\qquad \forall i=1,\ldots,N, \\
\max_{i=1,\ldots,N} l^T l_i &> 0, \qquad \forall|l|=1;
\end{align}
\end{subequations}
\item any $n$ vectors among $l_1, \dots, l_N$ are linearly independent.
\end{itemize}
Under Assumption \ref{assumption_command}, with $\bar{v}_c^d$ and $\bar{v}_c^{dd}$ arbitrarily small, by appropriately choosing the controller parameters $k_v$ and $k_p$, there exist positive constants $\bar{p}_0$ and $\bar{v}_0$ such that with the initial states satisfying
\begin{subequations}\label{theorem_initial_condition}
\begin{align}
|p_i(0)-p_o(0)-l_i(D+D_o)|&<\bar{p}_0~\text{for}~i=1,\ldots,N,\\
|v_o(0)| &\le \bar{v}_0,
\end{align}
\end{subequations}
the following properties hold:
\begin{itemize}
\item the position trajectories $(p_o(t),p_1(t),\ldots,p_N(t))$ are defined and the object's velocity $v_o(t)$ is bounded for all $t\geq 0$;
\item there exists a positive constant $\delta_v$ such that
\begin{align}
\limsup_{t\rightarrow\infty}|v_o(t)-v_c(t)|<\delta_v. \label{eq_theroem_results2}
\end{align}
\end{itemize}
\end{theorem}

See \cite{Davis-AJM-1954} for the concept of the positive span. The existence of $l_1,\ldots,l_N$ satisfying the conditions stated in Theorem \ref{theorem_main} has been justified by related studies; see, e.g., \cite{Wu-Liu-Niu-Jiang-RAL-2022, Wu-Liu-Egerstedt-Jiang-2023-TAC}. The proof of the theorem is postponed to Section \ref{section_proof}.

\begin{remark}
It should be noted that the design can be readily implemented in a distributed system setup by employing a distributed algorithm to solve the QP \eqref{eq_velocityloop_QP}. See, e.g., \cite{Nedic-Liu-ARCRAS-2018,Yang-Johansson-ARC-2019} for a tutorial on distributed optimization subject to inequality constraints.
\end{remark}

\section{Proof}
\label{section_proof}

The design in Section \ref{section_controller_design_new} basically forms the closed-loop system as a composition of two subsystems, which correspond to the velocity-tracking error $v_o-v_c$ of the object and the position-tracking errors $p_i-p_i^*$ of the mobile robots, respectively. In the proof, we first formulate the interconnections between the two subsystems, and then prove the achievement of the control objective by analyzing the interconnected system.

\subsection{Velocity-Tracking Control of the Object}

For the QP \eqref{eq_velocityloop_QP}, we first show that by choosing $\epsilon$ small enough, there exist a positive constant $\delta\in(0,1)$ such that,
\begin{align}
-k_f\sum^N_{i=1}l_is_i^*=(1-\delta)(-k_v(v_o-v_c)+\dot{v}_c)\label{eq_proof_f_condition}
\end{align}
holds for any $(v_o-v_c,\dot{v}_c)$. Since $(s_1^*,\ldots,s_N^*)$ is the solution of the QP, we have
\begin{align}
&\left|k_f\sum^N_{i=1}l_is_i^*-k_v(v_o-v_c)+\dot{v}_c\right|^2 \notag \\
&\le\left|k_f\sum^N_{i=1}l_is_i^*-k_v(v_o-v_c)+\dot{v}_c\right|^2 +\epsilon \sum^N_{i=1}|s_i^*|^2 \notag \\
&\le\left|k_f\sum^N_{i=1}l_is'_i-k_v(v_o-v_c)+\dot{v}_c\right|^2 +\epsilon \sum^N_{i=1}|s'_i|^2\label{eq_proof_solution_difference1}
\end{align}
holds for any $s_i'\geq 0$ for $i=1,\ldots,N$. Set
\begin{align}
(s'_1,\ldots,s'_N) = [l_{i_1},\ldots,l_{i_n}]^{-1} \frac{k_v(v_o-v_c)-\dot{v}_c}{k_f},\label{eq_proof_solution_differenceset}
\end{align}
where $l_{i_1},\ldots,l_{i_n}$ be $n$ vectors from $l_1,\ldots,l_N$ such that $(k_v(v_o-v_c)-\dot{v}_c)/k_f$ can be represented as a positive combination of $l_{i_1},\ldots,l_{i_n}$. This is achievable as $l_1,\ldots,l_N$ positively span $\mathbb{R}^n$. Then,
\begin{align}
&\left|k_f\sum^N_{i=1}l_is_i^*-k_v(v_o-v_c)+\dot{v}_c\right|^2\nonumber\\
&\leq \epsilon \sum^N_{i=1}|s'_i|^2\nonumber\\
&\leq\epsilon |[l_{i_1},\ldots,l_{i_n}]^{-1}|^2 \left| \frac{k_v(v_o-v_c)-\dot{v}_c}{k_f} \right|^2, \label{eq_proof_solution_difference2}
\end{align}
which verifies \eqref{eq_proof_f_condition}. The validity of $[l_{i_1},\ldots,l_{i_n}]^{-1}$ is guaranteed by the condition that any $n$ vectors among $l_1,\ldots,l_N$ are linearly independent, which is stated in Theorem \ref{theorem_main}.

With $p_i^*$ chosen as in \eqref{eq_velocityloop_virtualcontroller}, using \eqref{eq_proof_f_condition}, we have
\begin{align}
p_i^*-p_o=l_i(D+D_o-s_i^*) \label{eq_proof_relative_position}
\end{align}
for $i=1,\ldots,N$, and using \eqref{eq_def_acceleration}, \eqref{eq_proof_f_condition} and \eqref{eq_proof_relative_position}, we obtain
\begin{align}
&f(p_1^*-p_o,\ldots,p_N^*-p_o)\nonumber\\
&=-k_f\sum_{i=1}^{N}\max\{D+D_o-|p_i^*-p_o|,0\}\frac{p_i^*-p_o}{|p_i^*-p_o|}\nonumber\\
&=-k_f\sum_{i=1}^{N}l_is_i^*\nonumber\\
&=(1-\delta)(-k_v(v_o-v_c)+\dot{v}_c). \label{eq_proof_f}
\end{align}
Note that the second equality in \eqref{eq_proof_f} holds only if $s_i^*<D+D_o$ for all $i=1,\ldots,N$.

Recall \eqref{eq_positionloop_objectvelocitytrackingerror}. When $p_i\neq p_o$ for all $i=1,\ldots,N$, the velocity-tracking error satisfies
\begin{align}
\dot{v}_o-\dot{v}_c&=f(p_1-p_o,\ldots,p_N-p_o)-\dot{v}_c\nonumber\\
&=f(p_1^*-p_o,\ldots,p_N^*-p_o)-\dot{v}_c+\Delta_f\nonumber\\
&=-(1-\delta)k_v(v_o-v_c)-\delta\dot{v}_c+\Delta_f \label{eq_proof_dvo}
\end{align}
with
\begin{align}
\Delta_f &=f(p_1-p_o,\ldots,p_N-p_o)\notag \\
&~~~-f(p_1^*-p_o,\ldots,p_N^*-p_o). \label{eq_proof_Df_def}
\end{align}
From \eqref{eq_def_acceleration}, we can directly verify that the function $f$ is Lipschitz continuous with respect to $(p_1-p_o,\ldots,p_N-p_o)$, and there exists a constant $L_f$ such that
\begin{align}
|\Delta_f|\leq L_f\sum_{i=1}^N|p_i-p_i^*| \label{eq_proof_Df}
\end{align}
on any closed set in which $p_i-p_o\neq 0$ for $i=1,\ldots,N$.

\subsection{Position-Tracking Control of the Mobile Robots}

If the function $\phi_i$ defined by the QP \eqref{eq_velocityloop_QP} is Lipschitz continuous, then from \eqref{eq_plantmodel_integrator}, \eqref{eq_objectmodel_doubleintegrator_objectkinematics} and \eqref{eq_positioncontroller}, we have
\begin{align}
D^+(p_i-p_i^*)&=\dot{p}_i-\dot{p}_o+l_iD^+s_i^*\nonumber\\
&=v_i-v_o+l_iD^+s_i^* \nonumber\\
&=-k_p(p_i-p_i^*)+l_iD^+s_i^* \label{eq_proof_dp}
\end{align}
where 
\begin{align}
|D^+s_i^*|&\leq L_{\phi}|\dot{v}_o-\dot{v}_c|+L_{\phi}|\ddot{v}_c|\nonumber\\
&\leq L_{\phi}(k_v|v_o-v_c|+|\Delta_f|)+L_{\phi}|\ddot{v}_c|\nonumber\\
&\leq L_{\phi}(k_v|v_o-v_c|+L_f\sum_{i=1}^N|p_i-p_i^*|)\nonumber\\
&~~~+L_{\phi}|\ddot{v}_c|. \label{eq_proof_Dsi}
\end{align}

Now, we prove the Lipschitz continuity of $\phi_i$. For this purpose, we rewrite the QP \eqref{eq_velocityloop_QP} as 
\begin{subequations}\label{eq_velocityloop_QP_new}
\begin{align}
&s^* = \argmin_{s} \frac{1}{2} s^T R s + r^T(v_o - v_c, \dot{v}_c) s, \\
&\text{s.t.} \quad I_{N} s \ge 0_{N \times 1},
\end{align}
\end{subequations}
where $s^* = (s_1^*, \dots, s_N^*)$, $R = 2\epsilon I_{N \times N} + 2k_f^2 [l_1, \dots, l_N]^T [l_1, \dots, l_N]$, and $r(v_o - v_c, \dot{v}_c) = 2k_f [l_1, \dots, l_N]^T \big(-k_v (v_o - v_c) + \dot{v}_c\big)$. It can be directly verified that the QP \eqref{eq_velocityloop_QP_new} admits a unique solution for all $(v_o - v_c) \in \mathbb{R}^n$ and $\dot{v}_c \in \mathbb{R}^n$ (this is ensured by the fact that the feasible set of \eqref{eq_velocityloop_QP_new} is non-empty and convex and that $R$ is positive definite, as shown by the projection theorem in \cite[Proposition B.11]{Bertsekas-book-1997}); moreover, the eigenvalues of the constant matrices $R$ and $I_{N \times N}$ are positive. Then, applying \cite[Theorem 3.1]{Hager-SIAMControl-1979}, we can guarantee that the solution $s^*$ is Lipschitz continuous with respect to $r(v_o - v_c, \dot{v}_c)$. Finally, the Lipschitz continuity of $r$ guarantees that of $\phi_i$ with respect to $v_o - v_c$ and $\dot{v}_c$.

\subsection{Stability Property of the Closed-Loop System}

Denote $x_1 = v_o-v_c$ and $x_2 = [p_1-p_1^*; \ldots; p_N-p_N^*]$. Define 
\begin{align}
V_1(x_1) = |x_1|, \qquad V_2(x_2) = \sum^{N}_{i=1}|p_i-p^*_i|.
\end{align}
We use $D^+V_1(x_1)$ and $D^+V_2(x_2)$ to represent the dini derivatives of $V_1(x_1(t))$ and $V_2(x_2(t))$ with respect to $t$, respectively. Then,
\begin{align}
D^+V_1(x_1) 
&= \frac{x_1^T}{|x_1|}(-(1-\delta)k_vx_1-\delta\dot{v}_c+\Delta_f)\notag \\
&= -(1-\delta)k_vV_1(x_1) + \frac{x_1^T}{|x_1|}\Delta_f- \delta\frac{x_1^T}{|x_1|}\dot{v}_c \notag \\
&\le -(1-\delta)k_vV_1(x_1) + L_fV_2(x_2) + \delta \bar{v}^{d}_c, \label{eq_proof_V1}
\end{align}
and 
\begin{align}
D^+V_2(x_2) 
&= \sum^{N}_{i=1} \frac{(p_i-p^*_i)^T}{|p_i-p^*_i|}(-k_p(p_i-p_i^*)+l_iD^+s_i^*) \notag \\
&= -k_p V_2(x_2) + \sum^{N}_{i=1} \frac{(p_i-p^*_i)^T}{|p_i-p^*_i|}l_iD^+s_i^* \notag \\
&\le -k_p V_2(x_2) + \sum^{N}_{i=1} |D^+s_i^*| \notag \\
&\le -k_p V_2(x_2) + \sum^{N}_{i=1} L_{\phi}(k_vV_1(x_1) + L_fV_2(x_2))\notag\\
&~~~+L_{\phi}\bar{v}^{dd}_c \notag \\
&= (-k_p+NL_{\phi}L_f) V_2(x_2) + N L_{\phi}k_vV_1(x_1) \notag \\
&~~~+L_{\phi}\bar{v}^{dd}_c.\label{eq_proof_V2}
\end{align}
Thus,
\begin{align}
\begin{bmatrix} D^+V_1(x_1) \\ D^+V_2(x_2) \end{bmatrix} 
\le A \begin{bmatrix}V_1(x_1)\\ V_2(x_2)\end{bmatrix} 
+ \begin{bmatrix}\delta \bar{v}^{d}_c\\ L_{\phi}\bar{v}^{dd}_c\end{bmatrix}
\end{align}
with
\begin{align}
A = \begin{bmatrix}
-(1-\delta)k_v&  L_f \\
N L_{\phi}k_v & -k_p+NL_{\phi}L_f
\end{bmatrix}.
\end{align}

By choosing 
\begin{subequations} \label{eq_proof_condion}
\begin{align}
k_v &> 0, \\
k_p &> \frac{2-\delta}{1-\delta}N L_{\phi}L_f,
\end{align}
\end{subequations}
we can guarantee that the matrix $A$ is Hurwitz, and
\begin{align}
\frac{L_f}{(1-\delta)k_v} \cdot \frac{NL_{\phi}k_v}{k_p-NL_{\phi}L_f} < 1, \label{eq_proof_condion_smallgain}
\end{align}
fulfilling the nonlinear small-gain condition \cite{Jiang-Teel-Praly-MCSS-1994,Jiang-Mareels-Wang-Auto-1996}. 

Then, we illustrate the existence of the solution to the closed-loop system. In particular, since $\phi_i$ is Lipschitz continuous in $v_o - v_c$ and $\dot{v}_c$ on $\mathbb{R}^n$, and $f$ is Lipschitz continuous on any closed set in which $p_i-p_o\neq 0$ for $i=1,\ldots,N$. Using \cite[Theorem 54]{Sontag-book-98}, we have, for any initial positions satisfying \eqref{theorem_initial_condition} with $\bar{p}_0<D+D_o$, the closed-loop system admits a unique trajectory over some time interval starting at $t=0$. 

Based on the above results, we prove the main result by using the small-gain theorem. Following standard input-to-state stability arguments \cite{Sontag-2007}, under Assumption \ref{assumption_command}, with $\bar{v}_c^d$ and $\bar{v}_c^{dd}$ arbitrarily small, by appropriately choosing the controller parameters $k_v$ and $k_p$ to satisfy \eqref{eq_proof_condion_smallgain}, there exist positive constants $\bar{p}_0$ and $\bar{v}_0$ such that with the initial states satisfying \eqref{theorem_initial_condition}, we have that there exist positive constants $c_1$, $c_2$ and $c_3$ such that 
\begin{align}
|V_i(x_i(t))| \le \max\{&c_1e^{-c_2t}(|V_1(x_1(0))|+|V_2(x_2(0))|), \notag \\
&c_3|[\bar{v}^{d}_c;\bar{v}^{dd}_c]| \} \label{eq_theorem_Lyapunovs}
\end{align}
for $t\in[0,\infty)$ and $i=1,2$. Indeed, when $\bar{v}_c^d$, $\bar{v}_c^{dd}$, $\bar{p}_0$ and $\bar{v}_0$ are chosen sufficiently small, the trajectories $p_i(t)$ and $p_i^*(t)$ for $i=1,\ldots,N$ and $t\ge0$ remain in a region where $\Delta_f$ defined in \eqref{eq_proof_Df_def} satisfies inequality \eqref{eq_proof_Df} with $L_f$. 

Moreover, property \eqref{eq_theorem_Lyapunovs} directly guarantees the existence of $(p_o(t),p_1(t),\ldots,p_N(t))$ and the validity of \eqref{eq_theroem_results2}. In addition, combining \eqref{eq_theorem_Lyapunovs} with the boundedness of $v^c$ assumed in \eqref{assumption_command} yields the boundedness of $v_o$.

This ends the proof of Theorem \ref{theorem_main}.

\section{Numerical Simulation} \label{section_simulation}

In this section, we employ numerical simulations to verify the effectiveness of the proposed design.

In the simulation, we consider a multi-robot system described by \eqref{eq_plantmodel_integrator} with $n=2$ and $N=3$, and a transported object modeled by \eqref{eq_objectmodel_doubleintegrator} and \eqref{eq_def_acceleration} with $k_f=30$. The radii of the robots and the object are $D=0.2$ and $D_o=0.6$, respectively. 

For the controller consists of \eqref{eq_velocityloop_virtualcontroller}, \eqref{eq_velocityloop_QP} and \eqref{eq_positioncontroller}, we choose
\begin{align}
l_i = \left[\cos \frac{2\pi (i-1)}{N},  \sin \frac{2\pi (i-1)}{N}\right]^T,
\end{align}
and try the following two sets of parameters for comparison:
\begin{align}
\text{Parameter set 1:}\qquad k_v&=0.5, && k_p=1.0,\\
\text{Parameter set 2:}\qquad k_v&=0.5, && k_p=0.1.
\end{align}

The velocity-command signal $v_c$ is chosen as
\begin{align}
v_c(t) = -\left[\cos\frac{\pi t}{10}, \sin\frac{\pi t}{10}\right]^T,
\end{align}
which obviously satisfies Assumption \ref{assumption_command}.

For the numerical simulation, the initial states are chosen as follows:
\begin{subequations}
\begin{align}
p_1(0) &= [-7; 1], \quad p_2(0) = [-9; 1], \quad p_3(0) = [-9; -1], \\
p_o(0) &= [-8; 0], \quad v_o(0) = [0; 0].
\end{align}
\end{subequations}

\begin{figure}[ht]
\centering
\includegraphics[width=\linewidth]{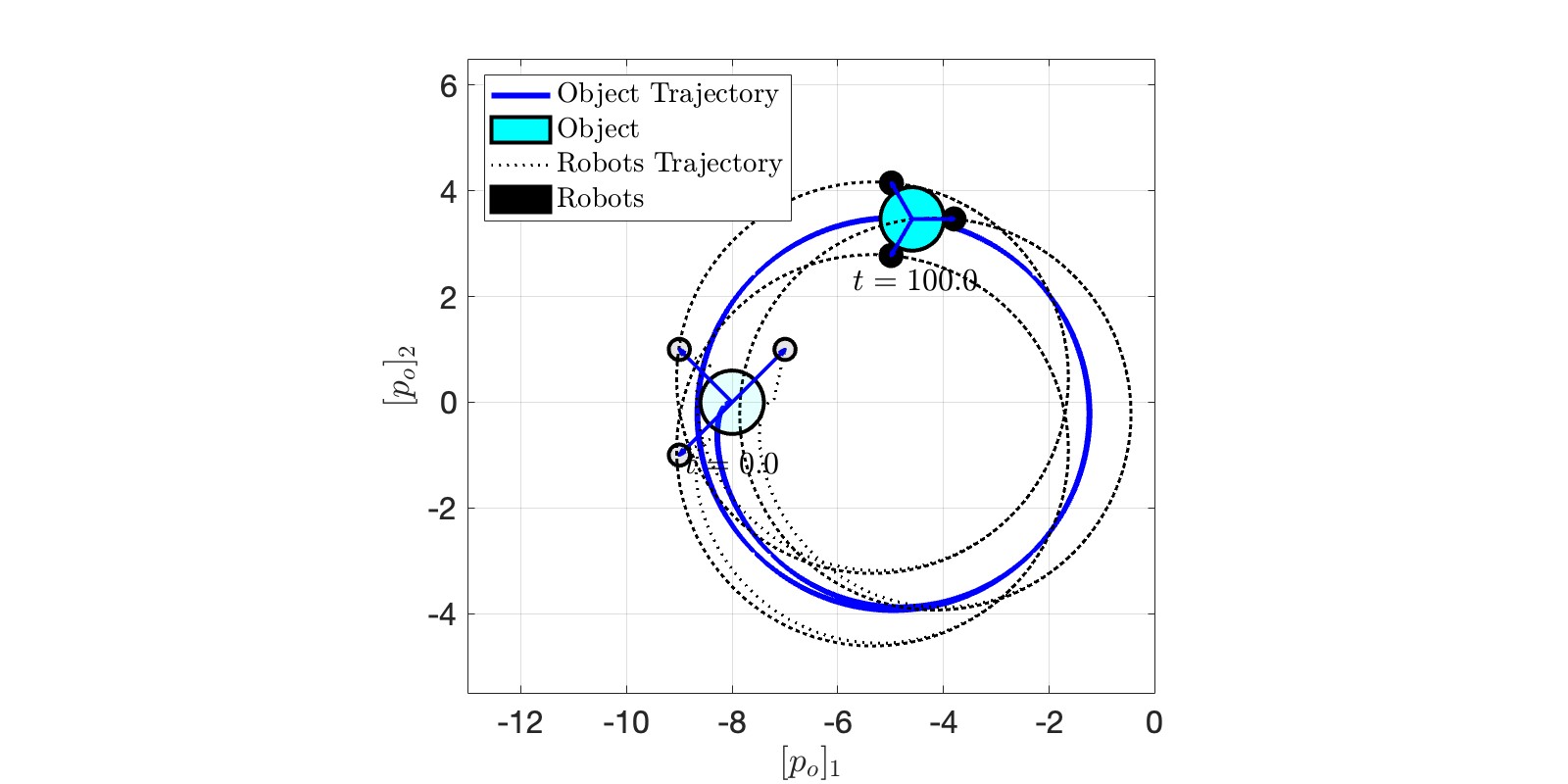}
\caption{The position trajectories of the robots and the object under Parameter Set 1.}
\label{figure_simulation_p}
\end{figure}

\begin{figure}[ht]
\centering
\includegraphics[width=\linewidth]{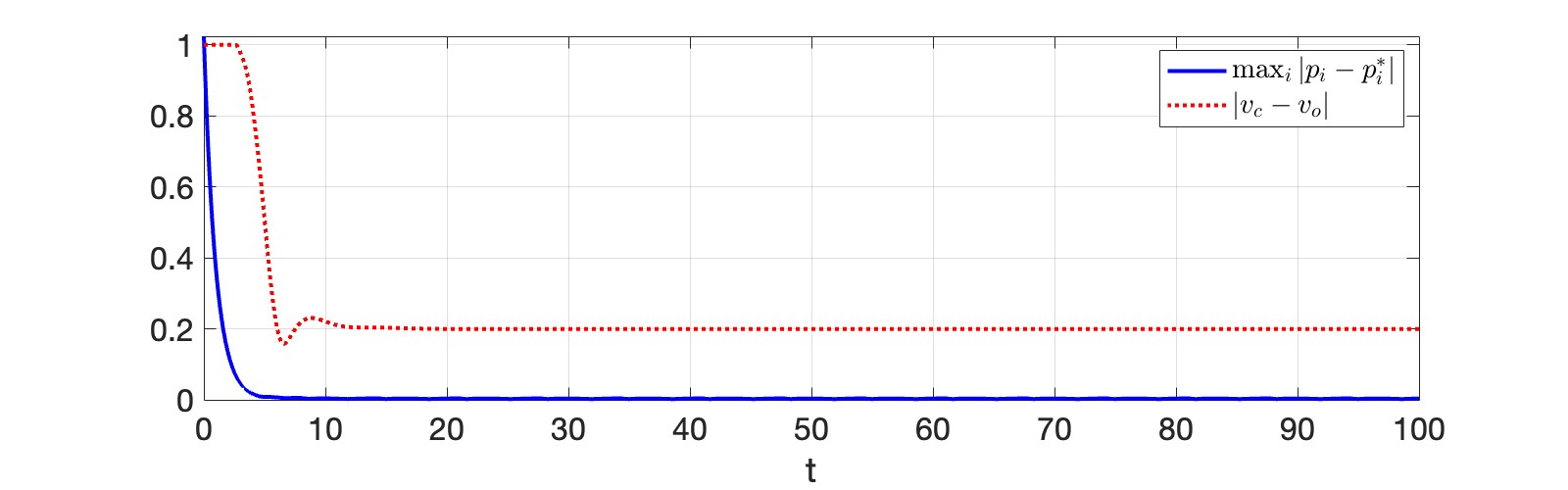}
\caption{The curves of $\max_i |p_i(t)-p^*_i(t)|$ and $|v_o(t)-v_c(t)|$.}
\label{figure_simulation_Vs}
\end{figure}

\begin{figure}[ht]
\centering
\includegraphics[width=\linewidth]{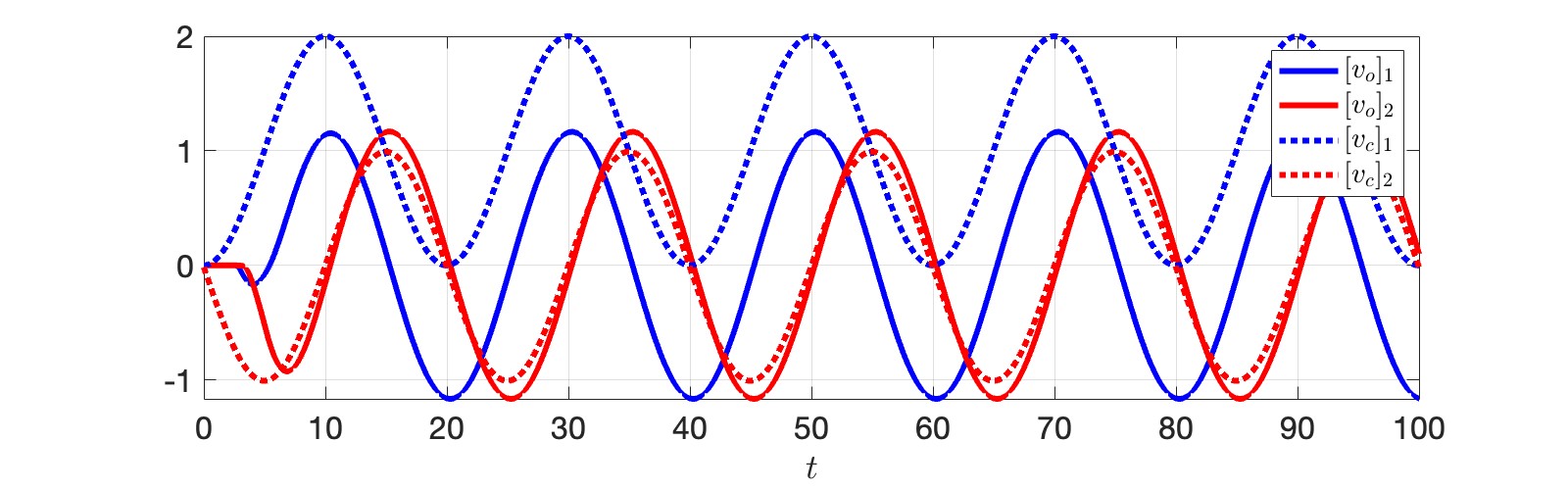}
\caption{The object velocity and the velocity-command signal.}
\label{figure_simulation_vo}
\end{figure}

\begin{figure}[ht]
\centering
\includegraphics[width=\linewidth]{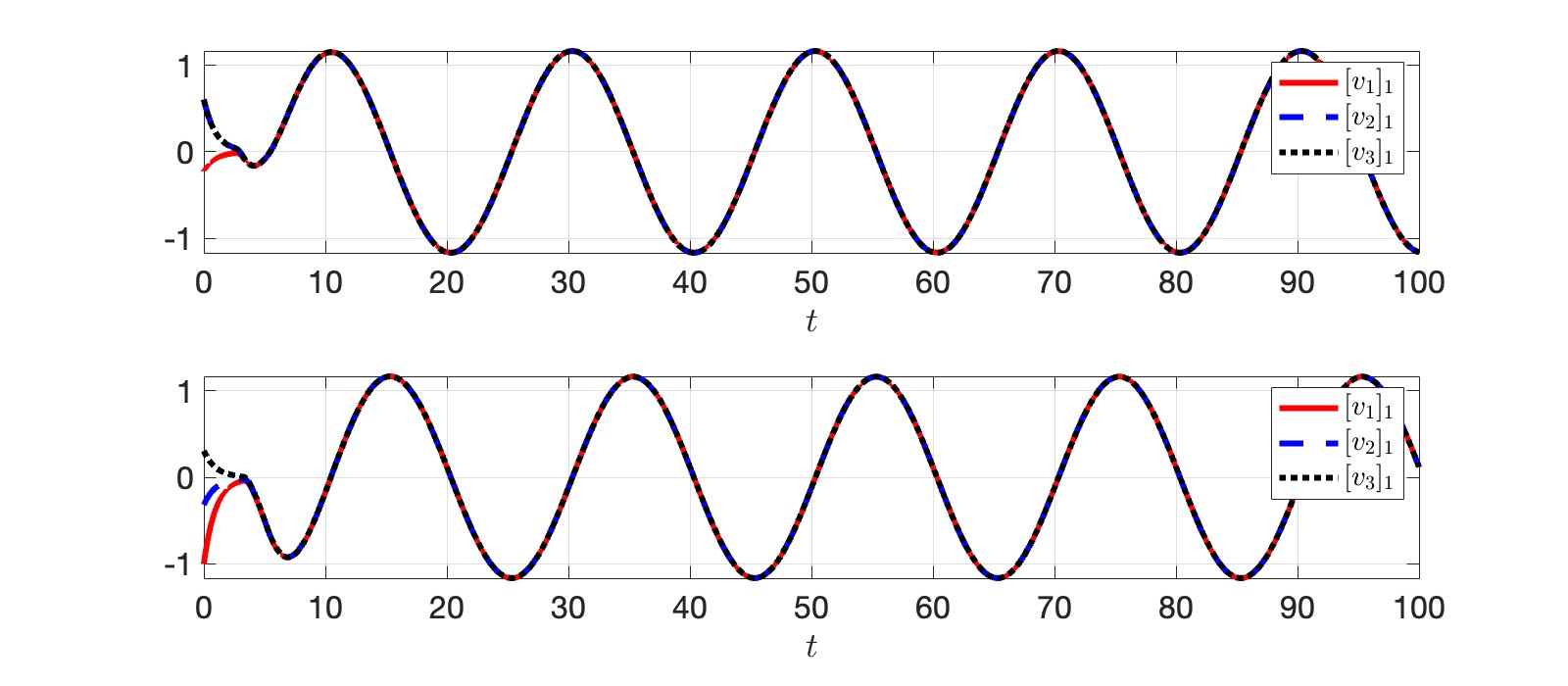}
\caption{The velocities (control inputs) of all the robots.}
\label{figure_simulation_v}
\end{figure}

\begin{figure}[ht]
\centering
\includegraphics[width=\linewidth]{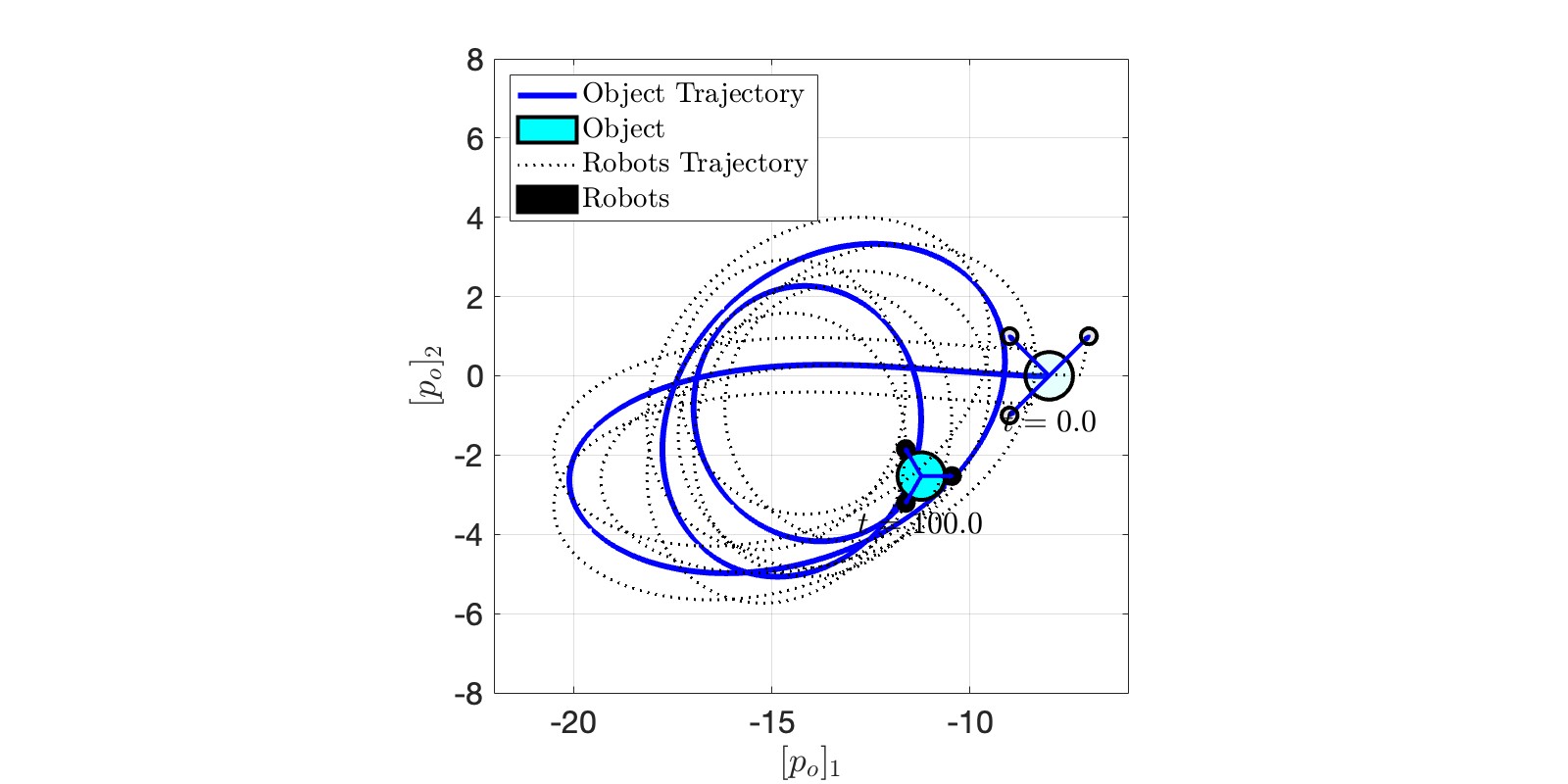}
\caption{The position trajectories of the robots and the object under Parameter Set 2.}
\label{figure_simulation_p2}
\end{figure}

Figures \ref{figure_simulation_p}, \ref{figure_simulation_Vs}, \ref{figure_simulation_vo} and \ref{figure_simulation_v} present the result for Parameter set 1, while Figure \ref{figure_simulation_p2} shows the result for Parameter set 2.

In particular, Figure \ref{figure_simulation_p} illustrates the position trajectory of the transported object, where the black small circles represent the robots and the cyan large circle denotes the object. It can be observed that, under the velocity-command signal $v_c$, the position trajectories of both the robots and the object form circular paths.

Figure \ref{figure_simulation_Vs} shows the curves of $\max_i |p_i(t) - p^*_i(t)|$ and $|v_o(t) - v_c(t)|$, which is consistent with the design in Section \ref{section_controller_design_new}. Figure \ref{figure_simulation_vo} illustrates that the object velocity $v_o(t)$ follows a similar trend to the velocity command signal $v_c(t)$. Figure \ref{figure_simulation_v} depicts the control input velocities of all robots.

Figure \ref{figure_simulation_p2} shows the positions of the robots and the object under Parameter set 2. It can be observed that the trajectories do not form regular circular paths as with Parameter set 1. 

\section{Conclusions} \label{section_conclusion}

In this paper, we have investigated the control problem of steering a group of spherical mobile robots to cooperatively transport a spherical object. To solve the problem, we have proposed a velocity-tracking controller based on QP, enabling the robots to cooperatively generate desired contact forces while minimizing the sum of the contact-force magnitudes, and then designed dedicated position-tracking controllers for the robots. The fundamental challenges caused by the uncertain dynamic coupling between multiple mobile robots and the object, as well as the requirement for real-time force distribution among the cooperating robots have been addressed by appropriately designing the QP. The achievement of the control objective is guaranteed by considering the closed-loop system as an interconnection of the velocity-controlled object and the position-controlled robots, and employing nonlinear small-gain techniques for stability analysis.

The results presented in this work lay the foundation for future research on more practical system setups, in which robot dynamics and nonholonomic constraints are explicitly considered, and the contact forces exhibit strongly nonlinear dependence on the relative positions between the robots and the object. Also, addressing safety constraints among the robots and the object remains a challenging problem. Advanced nonlinear control tools for interconnected systems, such as passivity-based methods and the nonlinear small-gain theorem, are expected to play an important role in overcoming these challenges.

\bibliographystyle{IEEEtran}
\bibliography{materials/underactuated_reference}

\end{document}